\documentclass [12pt]{article}

\usepackage {graphicx}
\usepackage{epsfig,amsmath,latexsym,amssymb}
\usepackage{bm}

\usepackage{booktabs}
\usepackage{multirow}
\begin{document}


\title{Pricing TARN Using a Finite Difference Method}

\author{Xiaolin Luo$^{1,\ast}$ and Pavel V.~Shevchenko$^{2}$}

\date{\footnotesize{Draft, this version 29 April 2013 }}

\maketitle

\begin{center}
\footnotesize { \textit{$^{1}$ CSIRO Mathematics, Informatics and Statistics, Sydney, Australia; e-mail: Xiaolin.Luo@csiro.au \\
$^{2}$ CSIRO Mathematics, Informatics and Statistics, Sydney,
Australia;
e-mail: Pavel.Shevchenko@csiro.au  \\
$^*$ Corresponding author} }
\end{center}

\begin{abstract}
\noindent Typically options with a path dependent payoff, such as
Target Accumulation Redemption Note (TARN), are evaluated by a Monte
Carlo method. This paper describes a finite difference scheme for
pricing a TARN option. Key steps in the proposed scheme involve
  tracking of multiple one-dimensional finite difference solutions, application of jump conditions at each cash flow exchange date, and
a cubic spline interpolation of results after each jump. Since a
finite difference scheme for TARN has significantly different
features from a typical finite difference scheme for options with a
path independent payoff, we give a step by step description on the
implementation of the scheme, which is not available in the
literature. The advantages of the proposed finite difference scheme
over the Monte Carlo method are illustrated by examples with three
different knockout types. In the case of constant or time dependent
volatility models (where Monte Carlo requires simulation at cash
flow dates only), the finite difference method can be faster by an
order of magnitude than the Monte Carlo method to achieve the same
accuracy in price. Finite difference method can be even more
efficient in comparison with Monte Carlo in the case of local
volatility model where Monte Carlo requires significantly larger
number of time steps. In terms of robust and accurate estimation of
Greeks, the advantage of the finite difference method will be even
more pronounced.

\vspace{1cm} \noindent \textbf{Keywords:} Target Accumulation
Redemption Note, option pricing, finite difference, Monte Carlo
\end{abstract}

\pagebreak

\section{Introduction}
\label{sec:introduction} Path dependent options have payoffs
depending on the trajectory followed by one or more of the
underlying processes. The most straightforward and easy to implement
numerical solution for pricing path-dependent options is based on
the Monte Carlo method. In the context of pricing path-dependent
options by solving partial differential equations (PDE), two
additional challenges may merge due to the presence of path
dependency. First, the dependency may introduce new dimensions to
the partial differential equation. Second, it may cause the
resulting equation much more difficult to solve because of the lack
of diffusion in the additional dimensions. For some detailed
discussions, see Tavella and Randall (2000)\nocite{Tavella2000},
Zvan et al (1998)\nocite{Zvan1998} and Wilmott
(2000b)\nocite{Wilmott2000b}.

The nature of the path-dependent option pricing problem largely
depends on whether we have a continuous or discrete sampling for the
path. In general, a continuous sampling model of path dependency
introduces additional convection terms in PDE, while for a
discretely sampled path-dependent option the convection terms are
replaced by jump conditions. There are many successful attempts in
pricing discretely sampled path dependent options by the PDE
approach using lattice based method (e.g. binomial and trinomial
trees used in Ritchken et al 1993\nocite{Ritchken1993}, Hull and
White 1993\nocite{Hull1993}, Barraquand and Pudet
1996\nocite{BarraquandPudet1996}, Forsyth et al
2002\nocite{Forsyth2002}), and similarly finite volume or finite
element method (Forsyth et al 1999\nocite{Forsyth1999}, Zvan et al
2001\nocite{Zvan2001}). Most of these studies consider Asian or
lookback options. Typically, a linear interpolation is adapted in
these methods in applying the jump conditions on the auxiliary
variable (e.g. path average of the underlying asset). The
convergence study by Forsyth et al (2002)\nocite{Forsyth2002} shows
that it is possible for an algorithm based on lattice method to be
non-convergent (or convergent to an incorrect answer) if the
interpolation scheme is selected inappropriately.

A Target Accumulation Redemption Note (TARN) provides a capped sum
of payments over a period with the possibility of early termination
(knockout) determined by the target level imposed on the accumulated
amount. A certain amount of payment (e.g. spot value minus the
strike) is made on a series of cash flow dates (referred to as
fixing dates) until the target level is breached. The payoff
function of a TARN is path dependent in that the payment on a fixing
date depends on the spot value of the asset as well as on the
accumulated payment amount up to the fixing date. Typically,
commercial software solutions for pricing a TARN are based on the
Monte Carlo method. This paper presents a finite difference scheme
as an alternative to the Monte Carlo method to evaluate TARN. The
focuses are on the step by step implementation of the finite
difference scheme, which is not readily available in the literature,
and on the comparison of performance of the proposed scheme relative
to the Monte Carlo. We are not aware of any finite difference scheme
published in the literature, although a general outline of PDE
approach to pricing TARN can be found in Piterbarg
(2004)\nocite{Piterbarg2004}.

Without losing generality, we assume the underlying asset is the
foreign exchange (FX) rate. The definitions of TARN options with
three different knockout types and some key notations are introduced
in Section 2. Foreign exchange rate models are described in Section
3. Finite difference scheme for TARN is presented in Section 4 and
numerical results for both the finite difference and Monte Carlo
methods are given in Section 5, before concluding by Section 6.

\section{TARN Payoff Definition}
\label{sec:mylabel1} There are different versions of TARN products
used in FX trading. For simplicity, here we consider one specific
form of TARN. The presented finite difference scheme can easily be
adapted to other more general forms of TARN as discussed in Section
\ref{secEX}. Denote the FX rate at time $t$ as $S(t)$ and other
notation as follows: $t_0$ is today's date; $K$ is the number of
fixing dates (cash flow dates); $t_1 ,t_2,\ldots,t_K$ are fixing
dates; $X$ is strike; $U$ is the target accrual level; $S(t_1
),S(t_2 ),\ldots,S(t_K )$ are FX rate values at fixing dates $t_1
,t_2 ,\ldots,t_K$; $A(t)$ is accumulated amount at time $t$; and all
amounts are per unit of notional foreign amount. On each fixing date
$t_k$, there is a cash flow payment

\begin{equation}
\label{eqCktilde} \widetilde{C_k}\equiv \beta (S(t_k)-X)\times
1_{\beta \times S(t_k) \ge \beta \times X},
\end{equation}
where $\beta $ is a strategy on foreign currency ($\beta = 1$
corresponds to buy and $\beta = - 1$ corresponds to sell), subject
to the target level $U$ is not breached by the accumulated amount
$A(t_k)$. If the target level $U$ is breached before or on the last
fixing date, denote $t_{\widetilde{K}}$ is the first fixing date
when the target is breached, i.e.
\begin{equation}
\label{eq3} \widetilde{K} = \min \{k:A(t_k) \ge U\},\;k =
1,2,\ldots,K \;.\end{equation} Otherwise, set $\widetilde{K}=K$. The
actual payment on the fixing date $t_k\le t_{\widetilde{K}}$ can be
written as
\begin{equation}
\label{eqCk} {C_k}(S(t_k),A(t_{k-1}))\equiv \widetilde{C}_k \times (
1_{A(t_{k-1})+ \widetilde{C}_k < U} + W_k \times 1_{A(t_{k-1})+
\widetilde{C}_k \geq U} ),
\end{equation}

\noindent and ${C_k}=0$ for $t_k> t_{\widetilde{K}}$. Here,
$A(t_{k-1})$ is the accumulation amount immediately after the fixing
date $t_{k-1}$, and $W_k$ is the weight depending on the type of the
knockout when the target level $U$
 breached. The accumulated
amount $A(t)$ is a piece-wise constant function $A(t)=A(t_{k-1})$,
$t_{k-1}\le t<t_k$ with

\begin{equation}\label{eq4}A(t_k) = A(t_{k - 1}) + {C_k}(S(t_k), A(t_{k - 1})
).
\end{equation}

There are three knockout types used in practice

\begin{itemize}
\item Full gain -- when the target is breached on a fixing date $t_k
$, the cash flow payment on that date is allowed. This essentially
permits the breach of the target once, and the total payment may
exceed the target for full gain knockout.

\item No gain -- when the target is breached, the entire payment on
that date is disallowed. The total payment will never reach the
target for no gain knockout.

\item Part gain -- when the target is breached on a fixing date $t_k $,
part of the payment on that date is allowed, such that the target is
met exactly.
\end{itemize}

\noindent Formally, it can be represented by the following
definition of the weight

\begin{equation}\label{eq5}W_k = \left\{ {{\begin{array}{*{20}l}
 {1,\quad \mbox{if knockout type} = \mbox{full gain};} \hfill \\
 {0,\quad \mbox{if knockout type} = \mbox{no gain};} \hfill \\
 {\frac{U - A(t_{k - 1}) }{\beta \times (S(t_k ) - X)},\quad\mbox{if knockout
type} = \mbox{part gain}.} \hfill \\
\end{array} }} \right. \end{equation}

The present value (discounted value) of the TARN payoff in domestic
currency for FX realization ${\rm {\bf S}} = (S(t_1 ),S(t_2
),\ldots,S(t_K ))$ is then

\begin{equation}
\label{eq1} P({\rm {\bf S}}) = \sum\limits_{k = 1}^{K} {\frac{C_k
\left(S(t_k), A(t_{k-1})\right) }{B_d (t_0 ,t_k )}}, \;\;\;
 A(t_0)=0,
\end{equation}

\noindent where $[B_d (t_0 ,t_k )]^{ - 1}$ is domestic discounting
factor from the fixing date $t_k $ to $t_0$.

Other forms of TARN used in trading include modifications of cash
flow payments (\ref{eqCktilde}) and accumulated amount rule
(\ref{eq4}). In the present study, the cash flow payment on each
fixing date is the same as the increment in the accumulated amount,
both are represented by
 $C_k(S(t_k),A(t_{k-1}))$.
 In other forms of TARN, the two quantities can differ, but
this should cause no additional difficulties for the finite
difference method presented here, as will be further discussed later
in Section \ref{secEX}.

\section{FX Model}
\label{sec:mylabel2} Under the standard no arbitrage option pricing
methodology, today's fair price of TARN is calculated as the
expectation of payoff (\ref{eq1}) under the risk neutral process.
Specifically, we consider the risk neutral process
\begin{equation}
\label{eq7} \frac{dS(t)}{S(t)}=(r_d-r_f)dt+\sigma dW_t,
\end{equation}

\noindent where $r_d $ and $r_f $ are domestic and foreign local
(instantaneous) interest rates, $\sigma $ is the local
(instantaneous) volatility and $W_t $ is the standard Brownian
motion. The expectation  can be calculated using Monte Carlo by
simulating risk neutral process (\ref{eq7}) many times and averaging
the payoff realizations; or by solving corresponding PDE via the
finite difference method. Here, the local interest rates can be
constant or functions of time $r_d = r_d (t)$, $r_f = r_f (t)$; and
volatility can be constant, function of time $\sigma = \sigma (t)$
or function of time and FX rate $\sigma = \sigma (S(t) ,t)$. The
last case corresponds to the local volatility model that can be
calibrated to match observed implied volatility surface; see e.g.
Wilmott (2000a)\nocite{Wilmott2000a}.

\section{Finite difference numerical scheme}
Let $V(S,t,A)$ be the value of TARN for spot rate $S$ and
accumulated amount $A$ at time $t$. Since the path-dependent
quantity $A$ is monitored discretely, there are no new diffusion
terms and the standard option pricing PDE is still valid between
fixing dates

\begin{equation}
\label{eq8} \frac{\partial V}{\partial
t}+\frac{1}{2}\sigma^2(S,t)S^2\frac{\partial V^2}{\partial S^2}
+(r_d(t)-r_f(t))S\frac{\partial V}{\partial S}-r_d(t)V=0.
\end{equation}

Typically, PDE solution for option pricing requires final conditions
(the payoff) and boundary conditions (e.g. at zero or at a barrier).
For discretely sampled path dependent options, additional jump
conditions apply. Unlike in the case of path independent options
where the payoff at expiry is known a priori and typically the final
condition is of the Dirichlet type with the value of the payoff, in
the case of the TARN option the final payoff is not known a priori.
The expiry time is simply the last fixing time, and the final payoff
depends on the path of the underlying up to the expiry time.
Immediately after the final payoff the option is worthless, and we
can set the final condition to zero at $T$
$$
V(S,T,A)=0,
$$
where $T=t_K$ is the  last monitoring time. Applying a proper jump
condition from  $T$ to  $T^- $ will give us a more informative final
condition at $T^- $, where $T^- $ is the time infinitesimally before
the last monitoring time $t_K=T$. Unfortunately, any single solution
of (\ref{eq8}) based on a given final condition at $T^- $ will not
lead to the correct answer to the TARN option pricing, even if we
know the final jump amount. We need multiple solutions to
(\ref{eq8}) with different final payoffs or jumps. Across any fixing
date, there is a discontinuous but predictable jump in the
accumulated amount. In such a case the no-arbitrage principle
dictates that there must be a proper jump condition imposed on the
path dependent option values. The jump value $C_k$ given by
(\ref{eq4}) is the cash flow to the TARN owner, thus

\begin{equation}
\label{eq9b} V(S ,t_k^ -,A(t_k^-)) =
V(S,t_k,A(t_k^-)+C_k(S,A(t_k^-))) + C_k(S,A(t_k^-)).
\end{equation}
Finally, the PDE solution will give us the today's TARN price
$V(S(t_0),t_0,0)$.

\subsection{Jump condition application}
Let us introduce an auxiliary finite grid $0 = A_1 < A_2 < A_3
\cdots < A_J = U$ to track the accumulated amount $A$, where $J$ is
the total number of nodes in the accumulated amount coordinate. The
upper limit $U$ is needed because the accumulated amount cannot
exceed the target $U$. For each $A_j $, we associate a continuous
finite difference solution to the one-dimensional PDE (\ref{eq8}).
For finite difference solution, at every jump we let $A$ to be one
of the grid points $A_j ,\;1 \le j \le J$. Since $A$ is always known
at each jump to be one of the fixed nodal point values, there is no
need to continuously track the actual evolution of the accumulated
amount $A$ during the entire finite difference solving process.

Denote finite difference grid points in the $S$ variable as $S_1
,S_2, \ldots,S_M$, where $M$ is the total number of nodes in the $S$
coordinate. For any $S=S_m$, $ m=1,\ldots, M$, substituting
$A(t_k^-)$ with $A_j$, $j=1,\ldots, J$ in (\ref{eq9b}) we get,

\begin{equation}
\label{eq9} V(S_m ,t_k^ -,A_j) = V(S_m ,t_k, A_j^+ ) + C_k(S_m,
A_j), \;\;\; A_j^+=A_j+C_k(S_m, A_j)\end{equation}

\noindent where $t_k^ - $ denotes the time infinitesimally before
the monitoring time $t_k $. In equation (\ref{eq9}), we have let the
accumulated amount before the k-th payment at $t=t_k^-$ to be one of
the grid point $A_j$. Equation (\ref{eq9}) describes a forward jump
from $t_k^-$ to $t_k$.

Because backward time marching is carried out for finite difference
solution of PDE (\ref{eq8}) associated with a fixed node point
$A_j$, intuitively the jump should be applied backwards from $t_k$
to $t_k^-$.  That is, in finite difference solution the value of $A$
at $t_k$ is known to be one of the grid point $A_j$, and after a
backward jump from $t_k$ to $t_k^-$ the value of $A$ changes from
$A_j$ to $A_j^-$. This backward jump can be expressed as

\begin{equation}
\label{eq10b} V(S_m ,t_k^ -,A_j^-) = V(S_m ,t_k, A_j) + C_k(S_m,
A_j^-),\;\;\; A_j^-=A_j-C_k(S_m, A_j^-)\end{equation}

In both (\ref{eq9}) and (\ref{eq10b}), $C_k(S,A)$ is calculated
according to (\ref{eqCk}).  Figure 1 illustrates the application of
jump condition (\ref{eq10b}).

\begin{figure}[htbp]
\begin{center}
\includegraphics[scale=0.8]{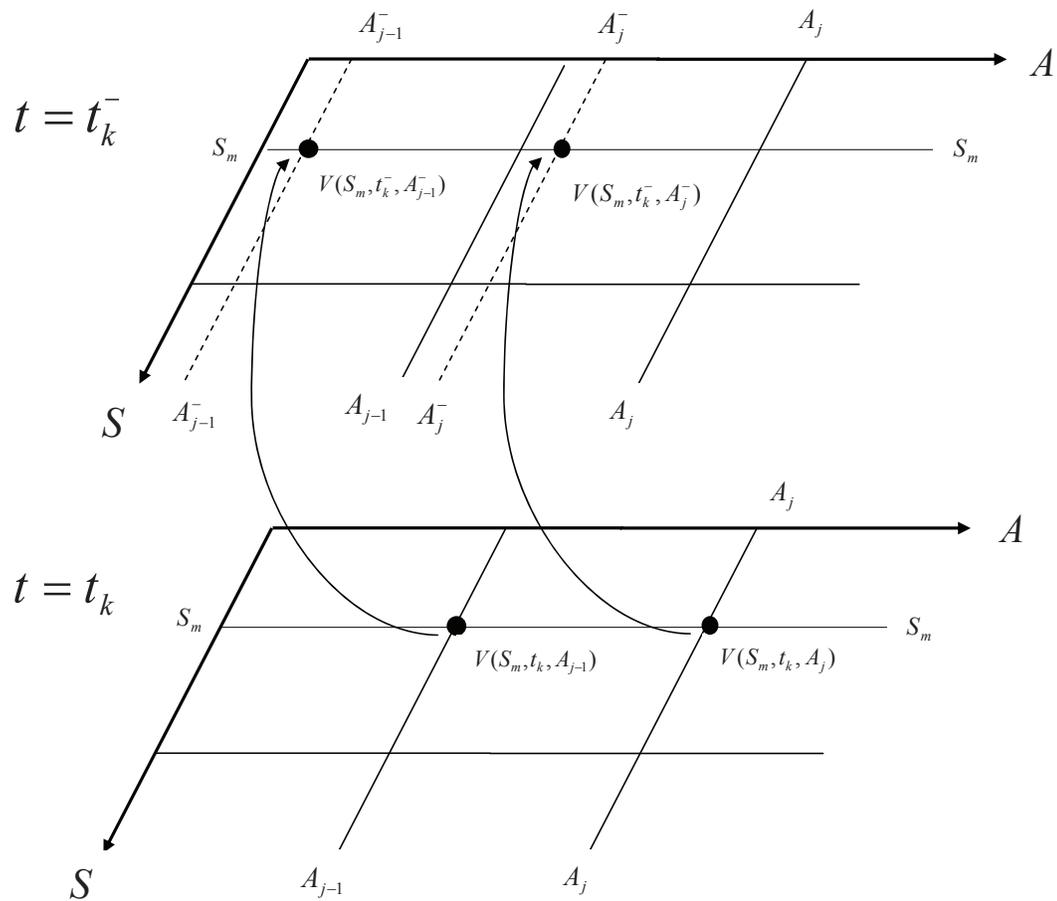}
\vspace{-8cm} \caption{Illustration of jump conditions applied to
finite difference grids.} \label{fig1}
\end{center}
\end{figure}

\subsection{Tracking finite difference solutions}
The idea is tracking $J$ finite difference solutions corresponding
to the $J$ grid points for the auxiliary variable, the accumulated
amount. For each fixed accumulated amount $A_j ,\;1 \le j \le J$, we
 start solving PDE by the finite difference scheme with the final condition
$V(S,T,A) = 0$ and a final jump condition from $t_K=T$ to
$t^-_K=T^-$. The implementation of this idea is not straightforward,
because at each sampling time, the jump condition (\ref{eq10b}) has
to be applied and the accumulated amount after each jump changes
accordingly and falls off the grid points $A_j ,\;1 \le j \le J$.
Not only the accumulated amount changes, the amount of changes
differ for different grid points in the underlying space. As shown
in (\ref{eq10b}), because $A_j^-$ is not a constant, the solution
$V(S_m ,t_k^- ,A_j^- )$ obtained after the jump does not correspond
to any grid point in the auxiliary variable space. Worse still, the
set of values $V(S_m ,t_n^ - ,A_j^- ),\;m = 1,\ldots,M$ does not
correspond to any continuous finite difference solution of the
one-dimensional pde -- it does not satisfy the PDE because the value
$A_j^-$ is scattered all over the place, not associated with any
unique value. This is because for the option value to satisfy the
one-dimensional pde, it requires a unique accumulated amount at any
time -- for consistency one cannot have different accumulated
amounts at the same time.

For the $M$ grid points in $S$ space, $V(S_m ,t_k^ - ,A^-_j )$, $m =
1,\ldots,M$ correspond to $M$ different scenarios of payoffs. Before
the jump, the $M$ values $V(S_m ,t_k ,A_j )$ are related to each
other through the PDE, because they are all associated to the same
accumulated amount $A_j $. The connection between the $M$ values is
broken after the jump.

\subsubsection{Reversal of the jump direction} Intuitively, jump
conditions should be applied through (\ref{eq10b}). That is, as the
backward marching is performed for each of the $J$ solutions
corresponding to $A_j ,\;1 \le j \le J$, at any crossing of sampling
time from $t_k $ to $t_k^ - $, $A_j $ jumps to $A_j^- $ and the
solution $V(S_m ,t_k ,A_j )$ jumps to $V(S_m ,t_k^ - ,A_j^- )$
according to (\ref{eq10b}). We can then interpolate from $V(S_m
,t_k^ - ,A_j^- )$ to obtain $V(S_m ,t_k^ - ,A_j )$ and continue time
marching backwards until next sampling date.

Unfortunately, the intuitive application of jump conditions as
described in the above paragraph is problematic in two important
ways.   First, from (\ref{eq10b}), it is possible to get a negative
value for $A^-_j $, which is invalid (meaningless) and out of the
range of the auxiliary variable space; Second, because $0 \le A_j
\le U$ and (\ref{eq10b}) for $A_j^-$ is a decreasing function, the
target $U$ will never be exceeded by any of the jump according to
(\ref{eq10b}), thus there is no way to apply the different knockout
conditions for the different knockout types as specified in
(\ref{eq5}). In other words, applying jump condition using
(\ref{eq10b})  cannot get the correct answers to any of the knockout
types of TARN. Essentially, applying (\ref{eq10b}) artificially
restricts the boundary for the auxiliary variable to be within the
target, instead of letting the underlying process breach the target.
Another minor issue is that the jump condition (\ref{eq10b}) is
implicit in $A^-_j$, i.e. strictly speaking the jump amount
$C_k(S_m, A^-_j)$ is not known before the backward jump.

The remedy to the above problems is actually quite simple -- we
reverse the direction of the jump. Jump condition (\ref{eq9b}) is
true for any values of the auxiliary variable in the range $0 \le A
\le U$, i.e. we do not have to use $A_j $, a grid point, on the
right hand side as in (\ref{eq10b}). Instead, we could use
(\ref{eq9}) to have the value of $A$ at $t_k^-$ to be one of the
gird point $A_j$. Now (\ref{eq9}) is explicit in $A_j$. Since grid
point $A_j $ satisfies $0 \le A_j \le U$ and $A_j^+ \ge A_j $, after
the jump from $A_j $ to $A_j^+ $, $A_j^+ $ will never be negative.
What is more, $A_j^+ $ value may now exceeds the target $U$,
allowing the knockout conditions to be imposed. The knockout
condition is implied in the calculation of $C_k(S_m, A_j)$ in
(\ref{eq9}), using equation (\ref{eqCk}). Specifically, we have

\begin{equation}
\label{eqCkb} {C_k}(S_m,A_j)\equiv \widetilde{C}_k \times ( 1_{A_j+
\widetilde{C}_k < U} + W_k \times 1_{A_j+ \widetilde{C}_k \geq U} ),
\end{equation}

Equation (\ref{eq9}) gives the desired solutions $V(S_m ,t_k^- ,A_j
)$ at the gird points $A_j ,\,j = 1,\ldots,J$, given $V(S_m ,t_k
,A_j^+ )$. Because we only have solution $V(S_m ,t_k ,A_j )$ upon
marching to time $t_k $, we need performing interpolation from
$V(S_m ,t_k ,A_j )$ to obtain $V(S_m ,t_k ,A_j^+ )$, for all
$m=1,2,\ldots , M$ and $j=1,2,\ldots , J$.

\subsubsection{Cubic Spline interpolation} For a fixed grid point
in spot $S = S_m $, there are $J$ values after the jumps
corresponding to the $J$ solutions associated with each of the $J$
gird points in the auxiliary variable. These values are given by
$V(S_m ,t_k ,A_j )$, $j = 1,\ldots,J$. We need to extract $J$ values
$V(S_m ,t_k ,A_j^ +)$ from $V(S_m ,t_k ,A_j )$, $j = 1,\ldots,J$ by
interpolating with respect to $A_j $. For a given $S_m $ this is a
one-dimensional interpolation in the accumulated amount space.

As shown in a convergence study by Forsyth et al
(2002)\nocite{Forsyth2002}, it is possible for a numerical algorithm
for discretely sampled path-dependent option pricing to be
non-convergent (or convergent to an incorrect answer) if the
interpolation scheme is selected inappropriately. All the previous
studies of numerical PDE solution for path dependent (Asian or
lookback options) used either a linear or a quadratic interpolation
in applying the jump conditions. In our experience a better choice
is the cubic spline interpolation (Press et al 1992\nocite{Pres92}).
This procedure assumes the $J$ values, $V(S_m ,t_k^ - ,A_j),j =
1,\ldots,J$, form a smooth function in the auxiliary variable space
and the cubic spline interpolation has a much higher order of
accuracy than linear or quadratic interpolation. The error of cubic
spline is $O(h^4)$ where $h$ is the size for the spacing of the
interpolating variable, assuming a uniform spacing. In our case $h =
\delta A = U / (J - 1)$. Natural boundary conditions are imposed at
the two ends $A_0 = 0$ and $A_J = U $, i.e. we assume zero second
derivative of the spline function at the two ends. For each fixed
spot $S_m $, a single tri-diagonal system of equations is solved
once for obtaining all the $J$ values $V(S_m ,t_k ,A_j^ +),\;j =
1,\ldots,J$.

If we perform the above interpolation for all the $M$ grid points in
spot $S$ and apply jump condition (\ref{eq9}), we will have $M\times
J$ new values $V(S_m ,t_k^ - ,A_j )$, $m = 1,\ldots,M$, $j =
1,\ldots,J$. For a fixed $j$, the $M$ new values $V(S_m ,t_k^ - ,A_j
)$ correspond to the PDE solution associated with grid point $A_j $.
Given $V(S_m ,t_k^ - ,A_j )$ for each fixed $A_j $, we can now
continue time marching backwards until the next sampling time. The
whole algorithm can be summarized as follows.

\begin{enumerate}

\item Apply zero final condition at $T=t_K$ for all the $J$ solutions
to equation (\ref{eq8}) corresponding to $A_j ,\;1 \le j \le J$.

\item Apply the jump condition (\ref{eq9}) to obtain $A_j^ + $ for each
of the $J$ solutions at each of the $M$ grid points in spot,
beginning with $k = K$ ($t_k = T)$ for the first jump.

\item Perform cubic spline interpolation from points $V(S_m ,t_k ,A_j
),j = 1,\ldots,J$ to new points $V(S_m ,t_k ,A_j^ + )$ by forming a
smooth function from the $J$ values $V(S_m ,t_k ,A_j )$ with each
spot grid point $S_m $.

\item Apply the jump condition (\ref{eq9}), i.e. calculate $V(S_m ,t_k^ -
,A_j )$ from $V(S_m ,t_k ,A_j^ + )$. The  knockout condition (a
boundary condition in variable $A$) is implied by the calculation of
 $C_k(S_m, A_j)$ using (\ref{eqCkb}).

\item Perform the finite difference time marching backwards for each
of the $J$ solutions $V(S ,t_k^ - ,A_j )$, $j = 1,\ldots,J$,
corresponding to the $J$ grid points in the auxiliary variable,
until a sampling time is encountered. This gives solution $V(S ,t_{k
- 1} ,A_j )$.

\item Repeat steps 2 to 5 until $k = 1$.

\item Take the single solution $V(S,t_1^- ,0)$ to do final time
marching until $t = t_0 $, and take $V(S,t_0 ,0)$ as the final
solution of the TARN option.

\end{enumerate}

As indicated in Step 1, at the final fixing time $T = t_K $, the
zero-value final condition is applied at $T$ and the following jump
condition is applied before taking any pde solving steps

\begin{equation}
\label{eq15} V(S_m ,T^ - ,A_j ) = 0 + C_k(S_m, A_j).\end{equation}

For each set of $V(S_m ,T^ - ,A_j )$ with fixed $j$, we begin
tracking a finite difference solution through backward time
marching.

In step 7, only a single solution is needed between the first
sampling time $t_1^ - $and the spot date -- there is no more need to
track all $J$ solutions, since there are no more jump conditions to
be applied. For good accuracy, we require that the current spot
value $S(t_0 )$ be one of the grid point in ${\rm {\bf S}} = (S_1
,S_2 ,\ldots,S_M )$.

\subsubsection{Extension to other TARN products \label{secEX}} The payoff structure with the three knockout
types of TARNs considered in this study is typical in FX trading,
but there are other types with different payoff structure and
knockout type. The extension of the present FD method to other TARN
types is straightforward. For example, suppose there are extra
payments $C_k^*$ at each fixing date $t_k$ and this extra payment
does not count in the knockout condition (\ref{eqCk}) but will also
get knockout by the same knockout condition  (\ref{eqCk}), i.e.

\begin{equation}
\label{eqCk2} C_k^*(S,A(t_{k-1}))=\widetilde{C_k^*} \times (
1_{A(t_{k-1})+ \widetilde{C_k} < U} + W_k \times 1_{A(t_{k-1})+
\widetilde{C_k} \geq U} ),
\end{equation}

\begin{equation}
\label{eqExtraPay} P({\rm {\bf S}}) = \sum\limits_{k = 1}^K {\frac{
 \left(C_k \left(S(t_k), A(t_{k-1})\right) +C_k^*\left(S(t_k), A(t_{k-1})\right) \right) }{B_d (t_0 ,t_k )}}, \;\;\;
 A(t_0)=0,
\end{equation}

\noindent where $\widetilde{C_k^*}$ is the extra payment when the
target is not breached. In this case, the only change in the finite
difference scheme is to replace the price jump condition (\ref{eq9})
with a new condition

\begin{equation}
\label{eqVjump2} V(S_m ,t_k^ - ,A_j ) = V(S_m ,t_k ,A_j^+ ) +
C_k(S_m, A_j) + C^*_k(S_m, A_j).\end{equation}

There is no any other change required in dealing with the auxiliary
variable $A_j$, since the extra payment does not contribute to the
 monitored accumulated amount $A$ and the knockout condition remains the same.

\subsection{Boundary condition} Typically a finite difference
solution is sought within a rectangular domain $(0 \le t \le T,\;0
\le S_{\min } \le S \le S_{\max } )$, where both $S_{\min } $ and
$S_{\max } $ are chosen to be sufficiently far away from the spot
price of the underlying asset, e.g. three standard deviations from
the spot. To insure an unique solution, boundary conditions are
required at $S_{\min } $ and $S_{\max } $. There are different ways
of imposing proper boundary conditions that are numerically
equivalent. A rather general and robust boundary condition at both
$S_{\min } $ and $S_{\max } $ is

$$
\frac{\partial ^2V}{\partial S^2}\left( {S_{\min } ,t} \right) = 0,\quad
\frac{\partial ^2V}{\partial S^2}\left( {S_{\max } ,t} \right) = 0,
$$

\noindent which is particularly useful because it is independent of
the contract being valued, provided the option has a payoff that is
at most linear in the underlying for small and large values of $S$
(almost all common contracts have this property). Other boundary
conditions work equally well. For example, for a call option, the
following boundary condition can be applied

$$ V\left( {S_{\min } ,t} \right) = 0,\quad \frac{\partial
V}{\partial S}\left( {S_{\max } ,t} \right) = 1,
$$

\noindent and for a put option we have

$$ V\left( {S_{\max } ,t} \right) = 0,\quad \frac{\partial
V}{\partial S}\left( {S_{\min } ,t} \right) = - 1.
$$

Some detailed discussions on various suitable boundary conditions
can be found in Wilmott (2000b).

\subsection{Log-transform} It is a common practice to re-write
equation (\ref{eq8}) in terms of $x = \ln (S)$ before finite
difference discretization:
\begin{equation}
\label{eq16} \frac{\partial V}{\partial
t}+\frac{1}{2}\sigma^2\frac{\partial V^2}{\partial
x^2}+\nu\frac{\partial V}{\partial x}-r_dV=0,
\end{equation}

\noindent where $\nu = r_d (t) - r_f (t) - \sigma ^2 / 2$. Equation
(\ref{eq16}) is slightly simpler than (\ref{eq8}), i.e. if
volatility and interest rates are constant, then coefficients of all
derivatives in (\ref{eq16}) are all constant.

\subsection{Discretization for uniform grid} Unlike barrier
options, pricing the discretely monitored TARN option can always
rely on uniform grids. This is because there are at most two
critical points to be `pinned' to grid points -- the spot and the
strike, provided we make the far boundaries flexible. Since the only
requirement for far boundaries is that they are sufficiently far
from spot, these boundaries can certainly be extended a bit further
to accommodate uniform grids with the two critical points (spot and
strike) pre-determined. When the spot and the strike are almost the
same, uniform grids tied to both the spot and the strike may have
too large a number of nodes, in this case we chose to tie the strike
only, and perform a one-off final interpolation to obtain the price
corresponding to the spot.

Denote the option price at time step $n$ and grid point $S_i $ as
$V_i^n $, $n = 0,1,2,\ldots,N$. For a uniform grid, $\delta x_i =
x_i - x_{i - 1} = \delta x$ is a constant, and we obtain the
following finite difference approximation with second order
accuracy

\begin{equation}
\label{eq17} \frac{\partial V}{\partial
x}(x_i,t_n)=\frac{V_{i+1}^n-V_{i-1}^n}{2\delta x}+O(\delta x^2),
\end{equation}
\begin{equation}
\label{eq18} \frac{\partial^2 V}{\partial
x^2}(x_i,t_n)=\frac{V_{i+1}^n-2V_i^n+V_{i-1}^n}{\delta x^2}+O(\delta
x^2).
\end{equation}

\vspace{0.5cm}

\noindent \textbf{The }$\theta - $\textbf{scheme} \\
Define the
following differential operator $F(V,x,\sigma ,\nu ,r_d )$
\begin{equation}
\label{eq19} F(V,x,\sigma ,\nu ,r_d
)\equiv\frac{1}{2}\sigma^2\frac{\partial V^2}{\partial
x^2}+\nu\frac{\partial V}{\partial x}-r_dV
\end{equation}

\noindent
and the associated finite difference operator $F_i^n $

\begin{equation}
\label{eq20} F^n_i\equiv\frac{1}{2}\sigma^2(x_i,t_n)\frac{\partial
V^2}{\partial x^2}(x_i,t_n)+\nu(x_i,t_n)\frac{\partial V}{\partial
x}(x_i,t_n)-r_d(t_n)V_i^n,
\end{equation}

\noindent
where the first and second derivatives are approximated by finite difference
as discussed above. Then the $\theta - $scheme can be expressed as

\begin{equation}
\label{eq21} \frac{V_{i}^{n+1}-V_{i}^n}{\Delta t}+\theta
F^{n+1}_i+(1-\theta)F_i^n=0,
\end{equation}

\noindent where $0 \le \theta \le 1$. Special values of $\theta =
0$, $\theta = 0.5$ and $\theta = 1$ correspond to fully explicit,
Crank-Nicholson and fully implicit scheme, respectively.

\section{Numerical examples}
Comparison of the finite difference and Monte Carlo methods is
performed in the case of basic model with constant volatility. In
this case, the number of time steps for Monte Carlo simulated paths
is the same as the number of fixing dates. In the case of basic or
term structure models, simulations between fixing dates are not
required because transition density between fixing dates is known in
closed form (it is just a lognormal density). For local volatility
model, simulations between fixing dates are required that will
increase computations proportionally to the number of time steps.

In the examples we consider all three types of knockout as described
in Section 2, each knockout type has four cases with four different
targets, so the total number of numerical examples is 12. The other
inputs common to all the examples are spot $S(0)=1.05$, strike
$X=1.0$, volatility $\sigma=0.2$, interest rates $r_d=r_f=0$, fixing
dates are every 30 days and we assume 20 fixing dates.

Results are summarised in Table \ref{Results_table}. As shown in
Table \ref{Results_table}, the computing time for Monte Carlo
estimates based on $N_{sim} = 200,000$ simulated paths is very close
to that for the finite difference method with mesh $500\times
100\times 500$ (500 points for spot, 100 points for accumulated
amount and 500 steps for time).

In Table \ref{Results_table}, the Monte Carlo standard error is
compared with estimated relative error of the finite difference
solution. Ideally, relative error should be computed as the relative
difference between numerical solution and the exact solution, for
both Monte Carlo and finite difference. Unfortunately in the case of
TARN options, closed form solution cannot be found except limiting
cases of one fixing date or very large target level. Nevertheless,
the standard error in Monte Carlo and the estimated relative error
in finite difference are both very good approximate to the exact
relative error. In the case of finite difference, we estimate the
relative error by using solution of the refined grids in spot,
accumulated amount spaces as well as in time. Specifically, we
double the number of grid cells in all three dimensions for the
refined calculation, i.e. using grids $1000\times 200\times 1000$
for spot, accumulated amount and time, and use this refined solution
in place of the exact solution in estimating the relative error. As
shown in the Appendix, because the $\theta - $scheme is second order
in accuracy in both spot space and time, and the cubic spline
interpolation in the accumulated amount is of the order $O(h^4)$,
using the solution of the refined grids in estimating the true
relative error of the coarser grids is valid and well justified.

\begin{table}[htbp]
{\footnotesize{\begin{tabular*}{1.0\textwidth}{cccccccc} \toprule
\textbf{target}& \textbf{MC}& \textbf{FD}& \textbf{diff {\%}}&
\textbf{stderr MC {\%}}& {\textbf{MC sec}}& \textbf{err FD {\%}}&
\textbf{FD sec} \\
\midrule
\multicolumn{8}{c}{\textbf{No gain} }  \\
\midrule 0.3& {0.1955}& {0.1955}& 0.0000{\%}& 0.10{\%}& 1.31&
0.045{\%}&
1.12 \\
 0.5& {0.3288}& {0.3286}& 0.0609{\%}& 0.10{\%}&
1.32& 0.001{\%}&
1.13 \\
 0.7& {0.4507}& {0.4505}& 0.0443{\%}& 0.10{\%}&
1.32& -0.018{\%}&
1.13 \\
 0.9& {0.5633}& {0.5633}& 0.0000{\%}& 0.10{\%}&
1.32& 0.015{\%}&
1.14 \\
\midrule
\multicolumn{8}{c}{\textbf{Part gain}}  \\
\midrule 0.3& {0.2446}& {0.2445}& 0.041{\%}& 0.08{\%}& 1.32&
0.016{\%}&
1.12 \\
 0.5& {0.3819}& {0.3818}& 0.0262{\%}& 0.09{\%}&
1.33& 0.005{\%}&
1.13 \\
 0.7& {0.5063}& {0.5061}& 0.0395{\%}& 0.10{\%}&
1.32& 0.038{\%}&
1.13 \\
 0.9& {0.6203}& {0.6200}& 0.0483{\%}& 0.10{\%}&
1.32& 0.010{\%}&
1.13 \\
\midrule
\multicolumn{8}{c}{\textbf{Full gain}}  \\
\midrule 0.3& {0.2979}& {0.2978}& 0.0336{\%}& 0.08{\%}& 1.32&
0.039{\%}&
1.12 \\
 0.5& {0.4389}& {0.4386}& 0.0684{\%}& 0.09{\%}&
1.33& 0.001{\%}&
1.12 \\
 0.7& {0.5646}& {0.5644}& 0.0354{\%}& 0.10{\%}&
1.33& 0.015{\%}&
1.13 \\
 0.9& {0.6792}& {0.6790}& 0.0295{\%}& 0.10{\%}&
1.32& 0.012{\%}&
1.13 \\
\bottomrule
\end{tabular*}
}} \caption{Finite Difference (FD) vs Monte Carlo (MC) results for
TARN price. The notional amount is one unit of foreign currency. The
column ``diff \%" shows the relative difference between results of
FD and MC. The computing time is for desktop with Intel Core i5-2400
@3.10GHz and 4 Gb RAM.} \label{Results_table}
\end{table}

As show in Table \ref{Results_table}, the accuracy of finite
difference solution is significantly better than that of the Monte
Carlo in all the 12 test cases. On average, the Monte Carlo standard
error is about 0.1{\%}, while the finite difference relative error
is about 0.02{\%}. That is, Monte Carlo relative error is five times
as large as finite difference relative error, thus on average Monte
Carlo computing time should increase by the factor of 25 to achieve
the same accuracy as finite difference because Monte Carlo standard
error is proportional to $1 / \sqrt {N_{sim} } $. Note that quoted
Monte Carlo relative error is computed from the standard error of
the estimate, i.e. it should be at least doubled for a more
realistic error estimate. To improve the accuracy of Monte Carlo
estimates, in our numerical example, we use the sum of vanilla
options with maturities at the fixing dates as a control variate
error reduction technique. Monte Carlo efficiency can  also be
improved by the use of other error reduction techniques such as
importance sampling described in Piterbarg
(2004)\nocite{Piterbarg2004} but it might be difficult to implement
this for more general models such as local volatility model and we
did not pursue this further.

These numerical results clearly demonstrate that the use of finite
difference will be even more beneficial (in terms of accuracy) in
the case of local volatility model where Monte Carlo method will
require simulations for extra time slices between fixing dates. We
expect that the impact in efficiency will be more pronounced in
calculation of Greeks where even small error in price such as
0.1{\%} may lead to 10-100{\%} error in second derivatives (e.g.
Gamma or Vanna).

\section{Conclusions}
\label{sec:conclusionsical} We have implemented a finite difference
scheme for evaluating TARN options. Numerical results show that
finite difference scheme is more efficient in pricing TARN than the
Monte Carlo counterpart, even for basic models where the volatility
is constant or piecewise constant between fixing dates. For a
surface model, the computing time in the Monte Carlo method will
increase in proportion to the number of time steps in the surface
model, while the finite difference scheme presented here remain
essentially the same in terms of computing time. In the numerical
examples only price was considered. It is expected that if the
Greeks are considered in the comparison between finite difference
and Monte Carlo, the advantage of finite difference will be much
more significant. Even a small error in price such as 0.1{\%} may
lead to a large error 10-100{\%} in second derivatives (e.g. Gamma
or Vanna). Thus pricing TARN and its Greeks by the proposed finite
difference scheme provides significant practical advantage over the
commonly used Monte Carlo method.

We have given very detailed descriptions of the numerical steps
required in the finite difference scheme, so that readers can easily
follow the procedures to implement their own, and re-produce the
result if desired. The TARN structure considered in this study is
simple. However, implementation of the finite difference method can
be easily extended to a more generalised accumulation rule and TARN
parameters varying across fixing dates.

\section{Appendix: Estimation of Numerical Error}
\label{sec:appendix}

Denote as follows: $V$ is the exact solution, $\widetilde {V}$ is
the numerical solution of the coarser grids (e.g. $500\times
100\times 500$ for spot, accumulated amount and time), $\widetilde
{V}^\ast $ is the numerical solution of the refined grids doubled in
each direction (e.g. $1000\times 200\times 1000)$ and $\delta =
\widetilde {V} - V$ and $\delta ^\ast = \widetilde {V}^\ast - V$are
the absolute numerical errors of the two grids, respectively. Then
the relative difference between the numerical solutions of the
coarse girds and refined grids is

$$ \widetilde {\varepsilon } = \left| {\frac{\widetilde {V} - \widetilde {V}^\ast
}{\widetilde {V}^\ast }} \right| = \left| {\frac{\delta - \delta
^\ast }{\widetilde {V}^\ast }} \right|,
$$

\noindent and the true relative difference between the numerical
solution of the coarse girds and the exact solution is

$$ \varepsilon = \left| {\frac{\widetilde {V} - V}{V}} \right| = \left|
{\frac{\delta }{V}} \right|
$$

It is easy to show $\widetilde {\varepsilon }$ is a very good
approximation of $\varepsilon $. Due to the second order accuracy in
both time and space, and fourth order accuracy in the accumulated
amount cubic spline interpolation, $\delta ^\ast $ can be estimated
as $\delta ^\ast \cong \pm 2^{ - 8}\delta $. Thus

$$ \widetilde {\varepsilon } = \left| {\frac{\delta - \delta ^\ast
}{\widetilde {V}^\ast }} \right| \cong \left| {\frac{\delta \mp 2^{
- 8}\delta }{V + 2^{ - 8}\delta }} \right| = \left| {\frac{\delta
}{V}} \right|\left| {\frac{256 \mp 1}{256 + \delta / V}} \right|
\cong \left| {\frac{\delta }{V}} \right|,
$$

\noindent where the last approximation sign is due to $\left|
{\frac{256 \mp 1}{256 + \delta / V}} \right| \cong 1$, assuming
$\left| \delta \right| < < V$. Depending on the signs of the
absolute errors, $\widetilde {\varepsilon }$ could be slightly
overestimating or slightly underestimating the true relative error
$\varepsilon $. Thus using relative error between solutions of
coarser grids and the refined grids (with number of grids in all
dimensions doubled) as an estimate of the true relative error is
well justified for a numerical scheme with second order accuracy.

\bibliography{bibliography}
\bibliographystyle{acm}

\end{document}